\documentclass[prl,aps,twocolumn,groupedaddress,showpacs,floatfix]{revtex4}

\usepackage{amsmath,amssymb,multirow,epsfig,bm}

\newcommand{\etal}{{\it et al.,\;}} 
\newcommand{\beq}{\begin{equation}}
\newcommand{\eeq}{\end{equation}}
\newcommand{\bea}{\begin{eqnarray}}
\newcommand{\eea}{\end{eqnarray}}

\newcommand{\veps}{\varepsilon}

\newcommand{\nn}{\nonumber}
\newcommand{\benn}{\begin{displaymath}}
\newcommand{\eenn}{\end{displaymath}}

\begin{document}

\title{\bf A dilute atomic Fermi system with a large positive scattering length  }

\author{ Aurel Bulgac}
\affiliation{Department of Physics, University of
Washington, Seattle, WA 98195--1560, USA}

\author{Paulo F. Bedaque}
\affiliation{Lawrence-Berkeley Laboratory,
1 Cyclotron Road,  MS 70R0319,  Berkeley, CA 94720--8169, USA}

\author{Antonio C. Fonseca}
\affiliation{Centro de F\'{\i}sica Nuclear da Universidade de Lisboa, Av.
  Prof. Gama Pinto 2, 1649-003,
  Lisboa, PORTUGAL}

\date{\today}

\begin{abstract}

We show that a dilute atomic Fermi system at sufficiently low temperatures,
can display fermionic superfluidity, even in the case of a repulsive atom-atom
interaction, when the scattering length is positive. The attraction leading to
the formation of Cooper pairs is provided by the exchange of Bogoliubov phonons
if a fraction of the atoms form a BEC of weakly bound molecules.

\end{abstract}



\maketitle


Ever since the pioneering experiment of DeMarco and Jin \cite{marco}
there has been an ongoing search for the superfluidity in a dilute
fermionic atomic system \cite{thomas}. A recent experimental result from
Jin's group \cite{two} seem to point to the possible creation of an
even more intriguing system, a dilute atomic Bose system, subsequently shown to
be in a BEC state\cite{three}, coexisting with a dilute atomic fermionic fluid,
perhaps even a superfluid itself as we will show here, both superfluids
composed from one single atomic 
species. Taking advantage of a Feshbach resonance between given
hyperfine states all these authors \cite{two,three}
have manipulated the scattering length and made it
positive. Starting with an almost equal mixture of these hyperfine
spin sub--states and a relatively small and negative scattering length,
by ramping the magnetic field at various rates, these authors have
created a mixture of weakly bound fermionic atoms and unbound fermions, all of
them interacting with positive scattering lengths, as we shall show here. The weakly 
bound state formed is bosonic in character, with a binding energy
typically larger than the temperature of the atomic cloud \cite{three}. This binding
energy is approximately given by $\hbar^2/ma^2$, when the scattering
length is large. We shall often refer to these weakly bound molecule as dimers.
The fraction of such
molecules depends on the rate at which the magnetic field is ramped
across the Feshbach resonance, ranging from essentially zero in the
sudden limit and up to about 80\% in the adiabatic limit.
At $T=0$ and in the adiabatic limit one would
expect a full conversion of all fermions into such weakly bound
molecules, the so called BCS $\rightarrow$ BEC crossover
\cite{leggett,nozieres,randeria,mohit,griffin}.  Unless the magnetic
field is very close to the Feshbach resonance, the size of these
molecules is smaller than the average inter particle separation.
If the weakly bound molecules form a
BEC state, then the induced fermion--fermion interaction, due to the
exchange of Bogoliubov bosonic sound waves, is attractive. Very close
to the Feshbach resonance when the atomic scattering length is much
larger than the effective range, the full fermion--fermion interaction
can then become weakly attractive. Under these circumstances the BEC may coexist
with a weakly coupled BCS superfluid. 

Over the last two decades many authors, starting with Leggett
\cite{leggett,nozieres,randeria,mohit,griffin,carlson,abyy}, studied
the BCS--BEC crossover in Fermi systems, when the pairing coupling
constant is varied from very small to relatively large values. If the
pairing coupling constant is so weak, that in vacuum two fermions do
not have a bound state, one can expect a relatively small pairing gap
at a finite fermion density, due to the Cooper instability
\cite{cooper}. When $k_F|a|\ll 1$ and $a<0$ the fermion kinetic
energy dominates over the interaction energy and the Fermi system has
a positive internal pressure. In the dilute limit (when $k_F|a|\ll 1$ and where
$k_F$ is the Fermi wave vector and $a<0$ the scattering length), the pairing
gap is given by 
\beq
\Delta \approx  \left ( \frac{2}{e} \right ) ^{7/3} \frac{\hbar^2k_F^2}{2m}
    \exp \left ( \frac{\pi}{2k_Fa}\right ). \label{eq:Delta}
\eeq
Even in the dilute limit the determination of the pre--exponential
factor requires the evaluation of many--body diagrams beyond the naive
expected mean--field effects alone \cite{gorkov,heiselberg}.  At least
theoretically, one can contemplate the following question: ``How do
the properties of a Fermi system evolve if one were to increase the
pairing coupling constant until two fermions in vacuum could form a
bound state?'' The arguments presented in
Refs. \cite{leggett,nozieres,randeria,mohit,griffin} lead us to believe
that the many fermion system will most likely undergo a smooth
crossover from a weak coupling BCS state to a BEC state, made of
relatively tightly bound fermion pairs. The prospects of observing
such a crossover were quite dim until the recent series of experiments
\cite{two,three}. While for both positive and negative values of
the scattering length when $k_F|a|\ll 1$ the properties of a many
fermion system can be described perturbatively, in the intermediate
region, when the scattering length is comparable or larger than the
average inter particle separation the entire system is in
non--perturbative regime.  The basic properties of a Fermi system with
$|a|=\infty$ have been refined recently in a quantum
Monte Carlo variational calculation \cite{carlson}.

In a Fermi--Bose mixture the interaction responsible for the formation
of Cooper pairs could be an induced interaction
\cite{heiselberg,bardeen,viverit,stoof,pethick}. A moving fermion can
excite a sound wave in the Bose gas, which will perturb at a
relatively large distance another fermion. This situation is similar
to the electron--phonon model of BCS superconductivity. 
The boson induced interaction between fermions is in
momentum--frequency representation given by
\beq
U_{fbf}(q,\omega)=    U_{fb}^2
\frac{2n_b\varepsilon_q}{\hbar^2\omega^2-
      \varepsilon_q(\varepsilon_q+2n_b U_{bb}) },
\label{eq:ind}
\eeq
where $U_{fb}$ is the fermion--boson coupling constant, $q$ and
$\omega$ are the wave vector and the frequency exchanged between the
two interacting fermions, $\varepsilon_q=\hbar^2q^2/2m_b$, $m_b$ is
the boson mass, $n_b$ the boson number density, $U_{bb}=4\pi\hbar^2a_{bb}/m_b$
and $a_{bb}$ the boson--boson scattering length. In the weak coupling limit the
two paired fermions exchange a small energy $\hbar \omega
={\cal{O}}(\varepsilon_F)$ and a large linear momentum $q={\cal{O}}(
k_F)$ and then the induced interaction is predominantly static and
attractive. It is straightforward to show that in coordinate representation the
induced interaction is of Yukawa  type \cite{heiselberg,bardeen,viverit,stoof,pethick}
\beq U_{fbf}(r)= -\frac{U_{fb}^2}{U_{bb}} \frac{1}{4\pi\xi_b^2r}
\exp\left (-\frac{r}{\xi_b}\right ), \eeq
where
\beq \xi_b=\frac{\hbar}{2m_bs_b}=\frac{a_{bb}}{\sqrt{16\pi
n_ba_{bb}^3}}, \quad s_b^2=\frac{n_{b}U_{bb}}{m_b}, \label{eq:xi} \eeq
$\xi_b$ is the healing or coherence length and $s_b$ is the speed of
the Bogoliubov sound waves in the Bose gas. It is notable that the
strength of the induced interaction in momentum representation at
$q=0$ is independent of the boson density. The radius $\xi_b$ of this
interaction could exceed significantly $a_{bb}$ in a very dilute Bose
gas, where $n_ba_{bb}^3\ll 1$, see Eq. (\ref{eq:xi}).

Using a very simple semiclassical criterion suggested by Calogero
\cite{calogero}, one can show that the strength of this induced
interaction is typically weak. Neglecting for the moment the role of
$U_{ff}$, Calogero's condition that the potential $U_{fbf}(r)$ can
sustain at least one two--particle bound state is
\beq \frac{2}{\pi} \int_0^\infty dr \sqrt{ \frac{ -m_fU_{fbf}(r) }{
\hbar^2 }} = \sqrt{ \frac{8}{\pi} \frac{ U_{fb}^2 m_f }{ U_{bb} 4\pi
\hbar^2 \xi_b } } \ge 1. \label{eq:calog} \eeq
This criterion is saturated in the case of a simple square--well
potential and thus cannot be improved. If one assumes that all
scattering lengths and masses are comparable in magnitude, it is easy
to see that this condition can hardly be satisfied in the dilute
limit, when the healing/coherence length is large ($\xi_b\gg a_{bb}$)
and the right hand side of Eq. (\ref{eq:calog}) is $\propto \sqrt{
a_{bb}/\xi_b} \ll 1$. We have thus naturally arrived at the conclusion
that the net effect of a repulsive fermion--fermion interaction
$U_{ff}$ and of the attractive boson induced fermion--fermion
interaction $U_{fbf}$ could lead to at most a weak coupling pairing
gap in the fermion sector. Even in the absence of $U_{ff}$ the induced
interaction can lead to a weak coupling BCS gap only.

An extremely interesting system to consider consists of fermionic
atoms mixed with diatomic molecules of the same atom species, as in
the recent experiment of Jin's group \cite{two}.  One can change the
fermion--fermion scattering length to a large and positive value
$a_{ff}:=a\gg r_0$ ($r_0$ is the radius of the interaction, which is
also typically of the order of the effective range) by means of a
Feshbach resonance.  For the sake of
simplicity of the argument we shall assume that there are equal
amounts of two identical fermionic spin subspecies, which we shall
formally identify with the two spin states of a spin 1/2.  In such a
system one can observe the BCS--BEC crossover predicted two decades
ago \cite{leggett,nozieres,randeria,mohit,griffin}. What was not appreciated
until recently \cite{abpfb} is that because not only $a_{ff}=a>0$, but also
$a_{fb}= 1.179\ a >0$ \cite{skorniakov,paulo} and $a_{bb} >0$,
one can manufacture a metastable mixed Fermi--Bose system composed of
such fermion atoms and weakly bound fermion pairs. The size of the
fermion pair is ${\cal{O}}(a)\gg r_0$ and a dilute system of such pairs alone
is metastable, since $a_{bb} = 0.60a >0 $ see below and Ref. \cite{dima} 

This is due to the act that the transitions between atoms and molecules is a
relatively slow process, see Refs. \cite{two,three} and in particular the
discussion of the latest experiment in Jin's group \cite{three}.  
In fact, at temperatures smaller than the binding energy
of a pair $T\ll \hbar^2/m_fa^2$, the pair--pair collision is mostly
elastic. Inelastic collisions with the formation a tightly bound state and one
or two fast atoms with kinetic energies of the order $\hbar^2/m_fr_0^2\gg
\hbar^2/m_fa^2\gg T$ is possible. The branching ratio for such a
channel is small ($\propto (r_0/a)^\alpha, \; \alpha{\approx 3}$) \cite{dima}. 
Moreover, since the final products will carry a large kinetic energy,
they will interact with the rest of the slow constituents with
cross sections of the order $r_0^2\ll a^2 \ll n^{-2/3}$, where $n$ is
either the fermion or the pair number density. A fast atom will not
interact with a pair with a cross section of the order of $a^2$, as
one might naively expect, for similar reasons why cosmic neutrinos do
not have a cross section of the order of the Earth radius squared when
they impinge on our planet.  Therefore, such process could not lead to
a noticeable heating of the system. Three--body recombination rates
can be suppressed by choosing appropriate low densities.  One should recognize
however that due to the different dependence on the scattering length $a$ 
of the three-body recombination and of the the rate of formation of tightly
bound molecules, the optimal conditions for the formation of an
atomic-molecular system correspond to such values of the scattering length for
which these rates are approximately equal.The lifetimes of such a mixed
atomic-molecular system could be estimated as ranging from $10^{-2}s$ to
several seconds, depending on the atom species, peak number density and 
specific value of the scattering length.

Th total energy density of such a system of spin 1/2--fermions is given by the
following expression ($m_f:=m, \; m_b:=2m\; na^3 \ll 1, \; n=n_f+2n_b$) \cite{abpfb}
\bea & & {\mathcal{E}}= \frac{3}{5}\frac{\hbar^2k_F^2}{2 m}n_f
+\frac{\pi\hbar^2 a}{m} n_f^2 \nn \\ & & + \frac{3.537 \pi\hbar^2
a}{m} n_fn_b +\frac{0.6\pi \hbar^2 a}{m}n_b^2 +\varepsilon_2 n_b ,
\label{eq:fermions} \eea
where $n_f=k_F^3/3\pi^2$, $k_F$ is the Fermi wave vector and
$\varepsilon_2=-\hbar^2/ma^2$ is the ground state energy of the pair.
The specific numerical values quoted here were obtained from a detailed
calculation of the atom-dimer and dimer-dimer scattering lengths described in
the next paragraph.There are well known corrections of higher order in
$na^3\ll 1$ to this expression, which we neglect here.  The chemical
potentials for the fermions and bosons in this case can be specified
independently. Chemical equilibrium between fermions and pairs can be
established only after a very long time, as at these small
temperatures $|\veps_2| =\hbar^2/ma^2\gg T$, see the discussion of recent results from
Jin's group \cite{three}.  The energy density in
Eq. (\ref{eq:fermions}) has to be supplemented with a contribution
arising from the weak coupling pairing correlations (condensation
energy), due to induced interactions, if there is a Cooper
instability. Since the Fermi subsystem is always in the weak coupling
BCS limit, this correction to the energy is small.

In the extreme limit when $a\gg r_0$, where $r_0$ is the effective
range appearing in the low energy expansion of the $s$--wave
scattering phase shift $k \cot \delta (k) = -\frac{1}{a}
+\frac{1}{2}r_0k^2 + \ldots,$ the fermion--pair scattering length is
$a_{fb}\approx 1.179\ a$, a result first established by Skorniakov and
Ter--Martirosian in 1957 \cite{skorniakov}. Since the value of the atom--atom
scattering length can be manipulated by means of the Feshbach
resonance, so can the ratio $r_0/a$, as $r_0$ (unlike $a$) should
have a small variation with the applied magnetic field in the narrow
region of the resonance. The dependence of $a_{fb}$ as a function of
$a=a_{ff}$ for fixed $r_0$ is determined, up to corrections of order
$(r_0/a)^3$, by a simple generalization of the Skorniakov and
Ter--Martirosian equation~\cite{skorniakov}. It turns out that the ratio
$a_{fb}/a$ varies by less than $1\%$ as $r_0$ varies from $r_0=0$
to $r_0=0.3a$ and the likelihood of Cooper pair formation is not much
affected by finite range effects. We have performed as well calculations 
of the dimer-dimer scattering length using a well established four-body
scattering formalism. The four-body equations we solve involve four identical
spin 1/2 fermions (two with spin up and two with spin down), 
interacting through a short range potential in $^1S_0$ alone. 
The formalism we use is identical to the one developed in Refs. \cite{af1}
to study the four-nucleon system at low energies and is based on the solution
of the Alt, Grassberger and Sandhas (AGS) equations \cite{ags} in the form that was
proposed by Fonseca and Shanley \cite{af2}. Since this method requires the solution of
the underlying two- and three-body T-matrix for all relevant channels, it is
appropriate to mention that we include in the calculation three-body
sub-amplitudes with total angular momenta ranging from $1/2^\pm$ to $9/2^\pm$, 
since we find that $P$-wave atom-dimer scattering contributes as much as 20 \% to the
dimer-dimer scattering length. The results of the calculation indicate that
$a_{bb} = 0.60a$. Calculations were performed using both a finite and a
zero range two-body interaction. In the case of a finite range we have varied
the ratio $a/r_0$ from ${\cal{O}}(1)$ to $\approx 2000$. The calculations of
Petrov {\it et al.} \cite{dima} of the atom-dimer and dimer-dimer scattering
lengths, based on a novel theoretical approach, whose validity we were unable
to assess though, agree with our result.  

\begin{figure}
\epsfxsize=7.0cm
\centerline{\epsffile{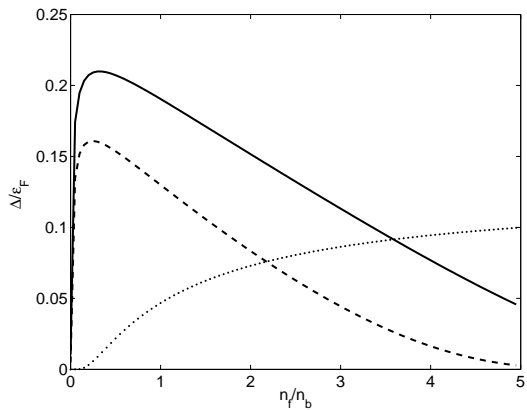}}
\caption{ \label{fig:fig1} The ratio $\Delta/\veps_F$ ($\veps_F=\hbar^2k_F^2/2m$) as a
function of $n_f/n_b$, for a fixed boson number density
$n_b=10^{13}\;cm^{-3}$ and $n_ba^3= 0.064$ (solid line) and
$n_ba^3=0.037$ (dashed line) respectively. The dots show the value of the gap
in the case of $p$-wave paring for $n_ba^3=0.064$.}  
\end{figure}

It is remarkable that the interaction properties of this mixed system of atoms
and weakly bound molecules can be described in terms of a single parameter, the
fermion--fermion scattering length $a$. From the previous analysis
\cite{heiselberg,bardeen,viverit,stoof,pethick} and the arguments
given above we know now that the tightly bound Cooper pairs, which
form a BEC, can lead to an additional attraction among the
fermions. The full fermion--fermion interaction in momentum space (and
zero frequency) is therefore
\beq 
{\mathrm{U}}_{ff}(q)= 
\frac{2\pi\hbar^2a}{m} \left [ 1 - \frac{5.213}{1+q^2\xi_b^2}\right ],
\eeq
where $\hbar \bm{q}$ is the momentum exchanged by two incoming fermions
with momenta $\hbar\bm{k}_{1,2}$. The $s$--wave pairing gap is determined by
the $l=0$ partial wave of this interaction (if attractive), when both 
fermion momenta are at the Fermi surface $(k_1=k_2=k_F$)
\bea \Delta& =& \left ( \frac{2}{e}\right )^{7/3}
 \frac{\hbar^2k_F^2}{2m} \nn \\ & \times &\exp \left [
\frac{\pi}{2k_Fa} \left (1 - 5.213 \frac{\ln
(1+4k_F^2\xi^2)}{4k_F^2\xi^2} \right )^{-1} \right ] , \eea
otherwise $\Delta\equiv 0$. The value of the pre-exponential factor requires
further analysis. The induced interaction has a relatively
large radius and typically in an extremely dilute system $2 k_F\xi= 1.127
(n_f/n_b)^{1/3}(n_ba^3)^{-1/6}\gg 1 $, if the densities of the fermion
and boson subsystems are comparable. Pairing in the $s$--wave is not
likely then, since in this case the $s$--wave full 
fermion--fermion interaction is repulsive. Still, for
appropriate values of the parameters s--wave pairing is possible, see
Fig. ~\ref{fig:fig1}.  For other values of the parameters $p$--wave pairing
should be considered \cite{efremov}, but the size of the $p$--wave gap
will typically be smaller.

The spectrum of sound waves in such a system has two branches. The
bosonic pairs will naturally have a Bogoliubov sound branch with the
speed $s_b$, see Eq. (\ref{eq:xi}), while in the fermion sector the
sound waves will have the speed \cite{anderson}, $s_f=
v_F/\sqrt{3}= \hbar(3\pi^2n_f)^{1/3}/\sqrt{3}m$,
unlike in a normal Fermi gas, where the speed of the Landau's zero
sound waves is larger then $v_F$ \cite{landau}. The ratio of these two
sound speeds could in principle have any value, depending on the
specific values of the scattering length and fermion and boson number
densities $s_f/s_b\propto ( n_f/n_b ) ^{1/3}/(n_ba^3)^{1/6}$.  For
dilute systems with comparable amounts of bosons and fermions the
slowest mode is the bosonic Bogoliubov sound wave.  There are
relatively small corrections to the values of $s_b$ and $s_f$, due to
interactions and also due to some mixing between the two types of
sound waves. We simply mention that for these systems (in both normal
and superfluid fermion phases) there is no long--wave instability of
the type suggested in Ref. \cite{meystre}. Even though the present analysis was
limited to the homogeneous case, the properties of a system in a trap can be
easily inferred and as, it was shown in Ref. \cite{trap}, an atom-molecule
mixture of the kind considered here acquires a complex and quite 
unexpected spatial structure.

Discussions with G.F. Bertsch, D.S. Petrov, G.V. Shlyapnikov, B. Spivak,
S. Stringari and L. Viverit  are greatly appreciated. AB and PFB acknowledge
DOE support and ACF was supported in part by grant POCTI/FNU/37280/2001.


\end{document}